\documentclass[preprint2]{emulateapj}
\usepackage{cancel}	
\usepackage{color}

\newcommand{\kms}{km\,s$^{-1}$}

\def\M{M$_{\odot}$}
\def\R{R$_{\odot}$}

\def\Ni{$^{56}$Ni}

 \def\Mej{$M_{\rm ej}$}
\def\Mcsm{$M_{\rm CSM}$}
\def\Mni{$M_{\rm Ni}$}

\def\Rs{$R_{\star}$}
\def\EM{$E_{\rm k}/M_{\rm ej}$}

\def\E{$E_{\rm k}$}

\shorttitle{LSQ14bdq: a double-peaked SLSN}
\shortauthors{M. Nicholl et al.}

\begin{document}


\title{LSQ14bdq: A Type Ic super-luminous supernova with a double-peaked light curve}


\author{M. Nicholl\altaffilmark{1}, 
S. J. Smartt\altaffilmark{1}, 
A. Jerkstrand\altaffilmark{1}, 
S. A. Sim\altaffilmark{1}, 
C. Inserra\altaffilmark{1}, 
J. P. Anderson\altaffilmark{2},
C. Baltay\altaffilmark{3},
S. Benetti\altaffilmark{4},
K. Chambers\altaffilmark{5},
T.-W. Chen\altaffilmark{1},
N. Elias-Rosa\altaffilmark{4},
U. Feindt\altaffilmark{6,7},
H. A. Flewelling\altaffilmark{5},
M. Fraser\altaffilmark{8},
A. Gal-Yam\altaffilmark{9},
L. Galbany\altaffilmark{10,11},
M. E. Huber\altaffilmark{5},
T. Kangas\altaffilmark{12},
E. Kankare\altaffilmark{1},
R. Kotak\altaffilmark{1},
T. Kr\"uhler\altaffilmark{2},
K. Maguire\altaffilmark{13},
R. McKinnon\altaffilmark{3},
D. Rabinowitz\altaffilmark{3},
S. Rostami\altaffilmark{3},
S. Schulze\altaffilmark{10,14},
K. W. Smith\altaffilmark{1},
M. Sullivan\altaffilmark{15},
J. L. Tonry\altaffilmark{5},
S. Valenti\altaffilmark{16,17},
D. R. Young\altaffilmark{1}
}

\altaffiltext{1}{Astrophysics Research Centre, School of Mathematics and Physics, Queens University
  Belfast, Belfast BT7 1NN, UK; mnicholl03@qub.ac.uk}
\altaffiltext{2}{European Southern Observatory, Alonso de Co\'rdova 3107, Vitacura, Santiago, Chile}
\altaffiltext{3}{Department of Physics, Yale University, New Haven, CT 06520-8121, USA}
\altaffiltext{4}{INAF - Osservatorio Astronomico di Padova, vicolo dell'Osservatorio 5, I-35122 Padova, Italy}
\altaffiltext{5}{Institute of Astronomy, University of Hawaii, 2680 Woodlawn Drive, Honolulu, Hawaii 96822, USA}
\altaffiltext{6}{Institut f\"ur Physik, Humboldt-Universit\"at zu Berlin, Newtonstr. 15, 12489 Berlin, Germany}
\altaffiltext{7}{Physikalisches Institut, Universit\"at Bonn, Nussallee 12, 53115 Bonn, Germany}
\altaffiltext{8}{Institute of Astronomy, University of Cambridge, Madingley Road, Cambridge, CB3 0HA, UK}
\altaffiltext{9}{Benoziyo Center for Astrophysics, Weizmann Institute of Science, Rehovot 76100, Israel}
\altaffiltext{10}{Millennium Institute of Astrophysics, Vicu\~{n}a Mackenna 4860, 7820436 Macul, Santiago, Chile}
\altaffiltext{11}{Departamento de Astronom\'ia, Universidad de Chile, Casilla 36-D, Santiago, Chile}
\altaffiltext{12}{Tuorla Observatory, Department of Physics and Astronomy, University of Turku, V\"ais\"al\"antie 20, 21500 Piikki\"o, Finland}
\altaffiltext{13}{European Southern Observatory, Karl-Schwarzschild-Strasse 2, 85748 Garching, Germany}
\altaffiltext{14}{Instituto de Astrof\'isica, Facultad de F\'isica, Pontificia Universidad Cat\'olica de Chile, Vicu\~{n}a Mackenna 4860, 7820436 Macul, Santiago, Chile}
\altaffiltext{15}{School of Physics and Astronomy, University of Southampton, Southampton, SO17 1BJ, UK}
\altaffiltext{16}{Department of Physics, University of California, Santa Barbara, Broida Hall, Mail Code 9530, Santa Barbara, CA 93106-9530, USA}
\altaffiltext{17}{Las Cumbres Observatory, Global Telescope Network, 6740 Cortona Drive Suite 102, Goleta, CA 93117, USA}

\begin{abstract}

We present data for LSQ14bdq, a hydrogen-poor super-luminous supernova (SLSN) discovered by the La Silla QUEST survey and classified by the Public ESO Spectroscopic Survey of Transient Objects. The spectrum and light curve are very similar to slow-declining SLSNe such as PTF12dam. However, detections within $\sim1$ day after explosion show a bright and relatively fast initial peak, lasting for $\sim15$ days, prior to the usual slow rise to maximum light. The broader, main peak can be fit with either central engine or circumstellar interaction models. We discuss the implications of the precursor peak in the context of these models. It is too bright and narrow to be explained as a normal \Ni-powered SN, and we suggest that interaction models may struggle to fit  the two peaks simultaneously. We propose that the initial peak may arise from the post-shock cooling of extended stellar material, and reheating by a central engine drives the second peak. In this picture, we show that an explosion energy of $\sim2\times10^{52}$\,erg and a progenitor radius of a few hundred solar radii would be required to power the early emission. The competing engine models involve rapidly spinning magnetars (neutron stars) or fall-back onto a central black hole. The prompt energy required may favour the black hole scenario. The bright initial peak may be difficult to reconcile with a compact Wolf-Rayet star as a progenitor, since the inferred energies and ejected masses become unphysical.

 \end{abstract}


\section{Introduction}\label{sec:intro}

Type Ic super-luminous supernovae (SLSNe) are hydrogen-poor explosions reaching absolute magnitudes $M_{\rm peak}<-21$ \citep{qui2011,gal2012,ins2013}. They are intrinsically rare \citep[less than $\sim$0.01\%~of the core-collapse population;][]{qui2013,mcc2015}, 
but their enormous electromagnetic output is observable at cosmological distances, and they show promise as standardisable candles \citep{ins2014}. However, the power source remains elusive. Viable models must account for the luminosity, blue colours \citep{qui2011},
spectroscopic evolution to resemble supernovae (SNe) Ic \citep{pas2010}, diverse light curves \citep{nic2015}, and low-metallicity environments \citep{nei2010,chen2013,lun2014,chen2014,lel2015}.

The slowest events, such as SN 2007bi \citep{gal2009,you2010} have been considered candidates for pair-instability supernovae \citep[PISNe; e.g.][]{heg2002}: complete thermonuclear disruptions of stellar cores with  $M_{\rm core}>65\,$\M. However, early observations of SLSNe apparently similar to SN 2007bi have shown rise-times and blue colours discrepant with numerical simulations \citep{kas2011,des2012,nic2013,mcc2014}.

Two further power sources have been proposed. One is a central engine, which could be the rotation of a millisecond pulsar with $B\sim10^{14}\,$G \citep{kas2010,woo2010}, or accretion by fallback if the SN forms a black hole \citep{dex2013}. The other is reprocessing of kinetic energy as the ejecta expand into a massive, extended circumstellar medium (CSM) \citep{woo2007,che2011,gin2012}. Both classes of models can fit SLSN light curves \citep{cha2013,nic2014}, but the spectra of SLSNe Ic do not show clear signs of CSM.

Here we present the discovery, light curve and analysis of the SLSN Ic, LSQ14bdq. Dense photometric sampling reveals an initial peak before the main light curve rise. We discuss physical interpretations of these data in the context of the competing models.

\section{Observations}\label{sec:obs}

\subsection{Discovery and spectroscopic classification}\label{sec:spec}

LSQ14bdq was discovered by La Silla QUEST \citep[LSQ;][]{balt2013}, at coordinates $\alpha = 10^{\rm h}01^{\rm m}41^{\rm s}.60$, $\delta = -12^{\rm o}22'13''.4$ (J2000.0), in images taken on April 5.1 UT (though earlier detections exist; Sect.~\ref{sec:phot}). It was classified by \citet{benitez2014}, as part of the Public ESO Spectroscopic Survey of Transient Objects \citep[PESSTO;][]{sma2014}, as a SLSN Ic. This spectrum (1500s) was taken on 2014 May 4.9 UT with the ESO 3.58m New Technology Telescope, using EFOSC2 and Grism\#13, and was followed by a longer exposure (2$\times$1800s) on the next night. A third spectrum was taken using Grism\#11 (2400s), on 2014 May 7.0 UT. These were reduced using the PESSTO pipeline, applying bias-subtraction, flat-fielding, wavelength and flux calibration and telluric correction \citep{sma2014}. The absolute fluxes were matched to contemporaneous photometry. PESSTO data are available from the ESO archive\footnote{see www.pessto.org} or WISeREP\footnote{http://www.weizmann.ac.il/astrophysics/wiserep/} \citep{yar2012}.

\begin{figure}
\includegraphics[width=\columnwidth,angle=0]{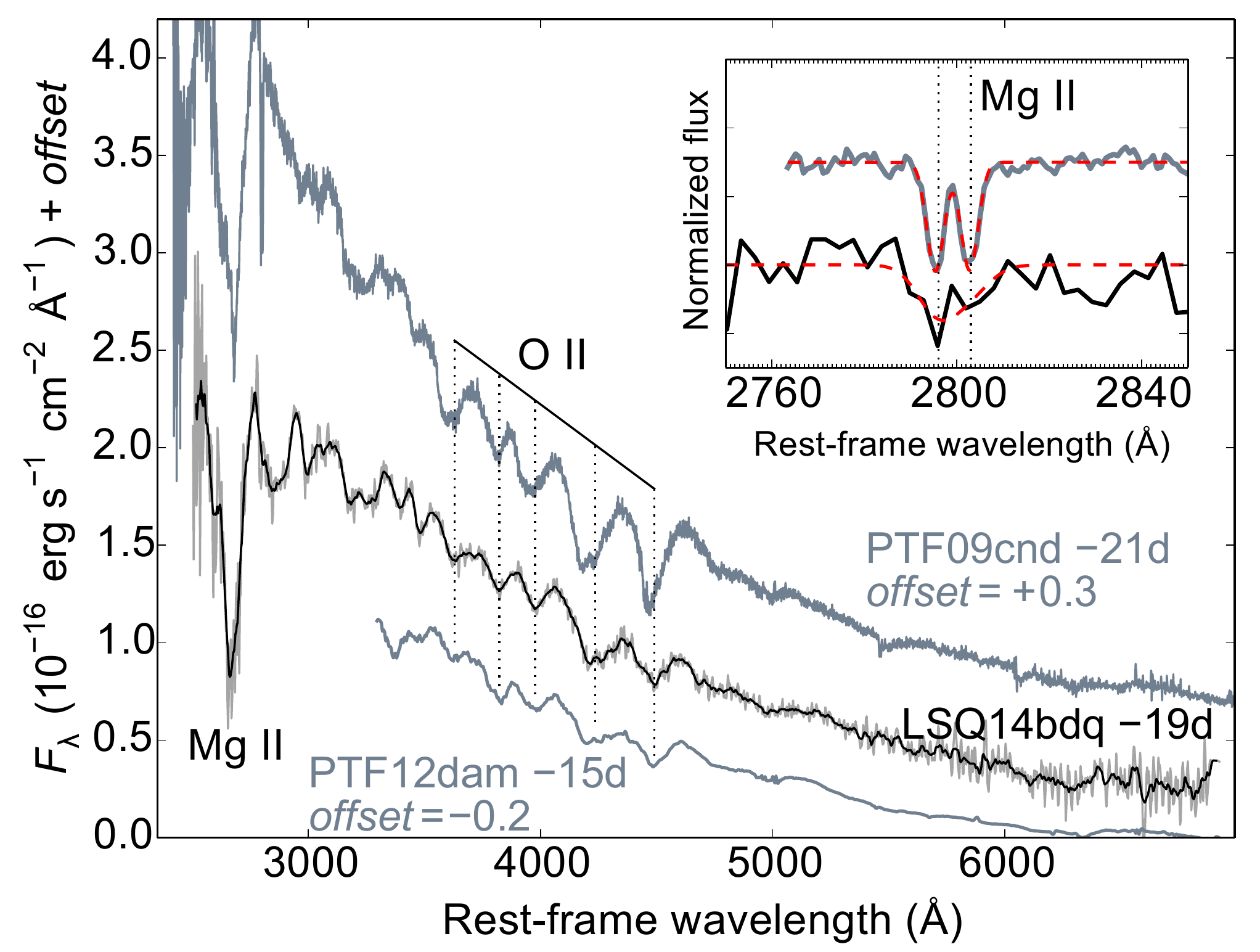}
\caption{The pre-maximum spectrum of LSQ14bdq (smoothed using
10-pixel moving average), compared to PTF12dam \citep{nic2013} and PTF09cnd \citep{qui2011}. The spectra are at similar phases from peak, and have been corrected for Galactic reddening \citep[$E(B-V)=0.056$;][]{schlaf2011}. The redshift is determined by fitting double-gaussian profiles (at instrumental resolution) to narrow Mg \textsc{II} absorption (inset, including PTF09cnd); the components are blended for the LSQ14bdq spectrum. Fluxes have been scaled to the  same luminosity distance as LSQ14bdq, and offsets added for presentation.
}
\label{fig:spec}
\end{figure}

Fig.~\ref{fig:spec} shows the spectrum summed over these three nights. The Grism\#11 spectrum (resolution 13.8\AA) shows interstellar Mg \textsc{II} $\lambda\lambda$2795.528, 2802.704 absorption, giving a redshift of $z=0.345$ from Gaussian fits. We estimate that the mean rest-frame phase of the combined spectrum is 19\,d before maximum brightness (Sect.~\ref{sec:phot}). Also shown are SLSNe Ic PTF12dam \citep{nic2013} and PTF09cnd \citep{qui2011}. 
The broad absorption features in LSQ14bdq are ubiquitous in such objects before maximum light. The O \textsc{II} lines are a defining feature of the class \citep{qui2011} and the deep Mg \textsc{II} absorption matches that in PTF09cnd and other objects with near-ultraviolet spectroscopy \citep[e.g.][]{chom2011}.

\subsection{Photometry}\label{sec:phot}

\begin{figure}
\includegraphics[width=\columnwidth,angle=0]{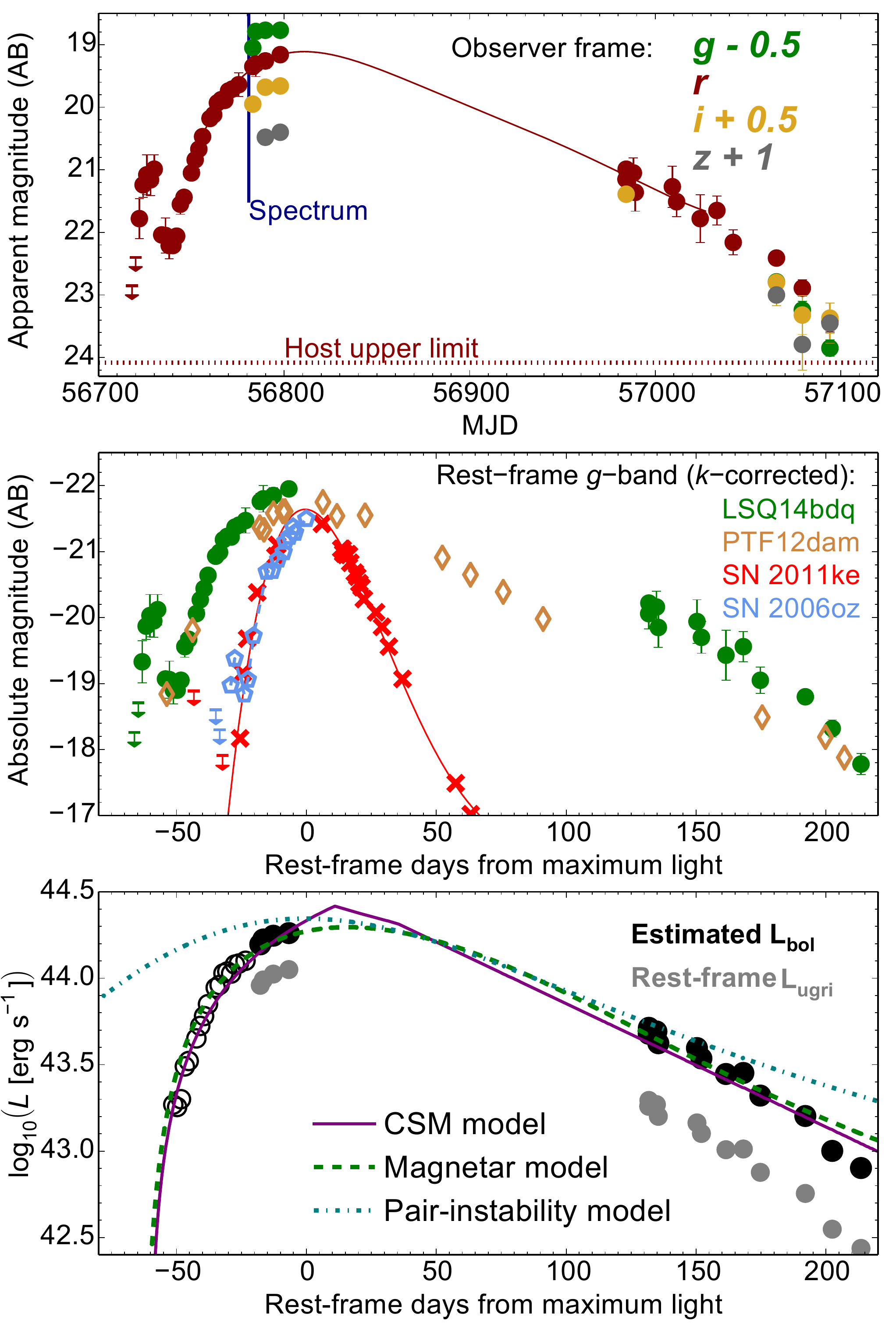}
\caption{The light curve of LSQ14bdq. Top: Observed photometry in $griz$. Polynomial fitting suggests a peak on MJD$\,=56807$, at $r=19.15$. Middle: Absolute light curve in rest-frame $g$-band, after $K$-correction and de-reddening, and resemblance to PTF12dam \citep{nic2013}. SLSNe 2011ke \citep{ins2013} and 2006oz \citep{lel2012} have good coverage during the rising phase. Bottom: The main peak of the bolometric light curve can be fit by magnetar and interaction models (see Sect.~\ref{sec:bol}). The rise of the PISN model is too slow to match the observations. Empty circles indicate points that are estimated from single-filter photometry (see Sect.~\ref{sec:bol}). 
}
\label{fig:lc}
\end{figure}


The observed light curve of LSQ14bdq is shown in Fig.~\ref{fig:lc}. The rise from 2014 March 22.1 UT was measured with the automated LSQ pipeline \citep{balt2013}, employing point-spread function (PSF) fitting forced photometry, and are calibrated to SDSS $r$ (AB system). Inspection of pre-discovery LSQ data showed clear variable flux at the SN position \emph{before} this first pipeline detection. Applying manual PSF photometry (using \textsc{SNOoPY}\footnote{http://sngroup.oapd.inaf.it/snoopy.html}) revealed that this was an early peak prior to the main rise. Non-detections (Table \ref{tab:phot}) suggest the explosion occurred on MJD\,$=56721\pm1$, assuming a smooth rise to the first peak. A stack of all LSQ images from 2012-2013 shows no host galaxy to a limiting magnitude of $r=24.1$, hence image subtraction is unimportant.

We obtained multicolour imaging using EFOSC2, the 2.0-m Liverpool Telescope and Las Cumbres Observatory Global Telescope 1-m network. PSF magnitudes were calibrated using a sequence of local field stars, themselves calibrated against standard fields on photometric nights. LSQ14bdq set for the season just before reaching maximum brightness. Subsequent data are a combination of PESSTO, LSQ, Pan-STARRS1 \citep[PS1;][]{mag2013} and GROND \citep{gre2008} images. The PS1 data were taken in the $w_{\rm P1}$-band \citep[effectively a $gri$ composite;][]{ton2012} from the Pan-STARRS NEO Science Consortium survey \citep{hub2015}, and also calibrated to $r$.

We converted observed $r$-band magnitudes to rest-frame $g$-band for comparison with other SLSNe (at this redshift, observed $g,r,i,z$ are similar to rest-frame $u,g,r,i$). The $K$-correction before peak is calculated from synthetic photometry on the LSQ14bdq spectrum ($K_{r \rightarrow g}\simeq-0.3$). As we only have one spectrum, our post-peak data use a $K$-correction calculated in a similar manner, but from the spectrum of PTF12dam at $+171$\,d \citep{nic2013}. LSQ14bdq has a broad light curve like PTF12dam. A polynomial fit suggests $M_g=-21.96$ at maximum light. The initial peak ($M_g=-20.01$) is much faster, with a rise of 5 days and a total width of $\sim15$ days. 

\citet{lel2012} presented pre-rise data for SN 2006oz with multi-colour detections. Fig.~\ref{fig:nickel} shows that the two are qualitatively similar. The rise to peak is not as pronounced as for LSQ14bdq, although time-sampling was sparser. \citet{lel2012} suggested a luminosity plateau powered by oxygen-recombination in extended CSM. Of the other SLSNe with strict explosion constraints, SN 2011ke \citep{ins2013} shows a similar rise to SN 2006oz, but non-detections shortly before discovery limit any early peak to be fainter than those observed for LSQ14bdq and SN 2006oz.


\section{Analysis}\label{sec:anal}

\subsection{Bolometric light curve and main peak}\label{sec:bol}

We constructed the bolometric light curve of LSQ14bdq in two steps. First we integrated the flux in the rest-frame $u$- to $i$-bands, following the procedure described in \citet{ins2013}. Before $-20$\,d, we caution that points are derived using the earliest available colour information. To estimate the full bolometric light curve, we take the fractional flux outside of this range to be the same as for PTF12dam \citep{nic2013,chen2014}. This seems reasonable, at least in the late-time NIR, from the $J$-band detection at 192 days (Table \ref{tab:phot}). Both the observed $ugri$-pseudobolometric and the estimated full bolometric light curve are shown in Fig.~\ref{fig:lc}.

We can reproduce the main peak with models powered by a central engine (we take a magnetar as representative) or by ejecta-CSM interaction. For details of the models, see \citet{ins2013,cha2012,nic2014}. The magnetar fit has magnetic field $B=0.6 \times 10^{14}\,$G, spin period $P=1.7\,$ms and diffusion time $\tau_m = 90$\,d. We have fixed the time of explosion to coincide with the first detection, on the precursor peak. The CSM model has ejected mass \Mej$\,=30.0\,$\M, CSM mass \Mcsm$\,=16.0\,$\M ~(assuming $\kappa=0.2\,{\rm cm}^2\,{\rm g}^{-1}$), density $\rho_{\rm CSM} = 3.0 \times 10^{-13}\,{\rm g}\,{\rm cm}^{-3}$ and explosion energy $E_{\rm k}=5.0 \times 10^{51}\,$erg.

We also compare to the brightest PISN model of \citet{kas2011} (a 130\,\M~bare helium core), which reproduces the peak luminosity. However, the rise-time is discrepant: the well-constrained main rise of LSQ14bdq lasts for 50 days, whereas the PISN model rises for over 100 days, and declines more slowly than LSQ14bdq. We therefore reach the same conclusion as \citet{nic2013} and \citet{mcc2014}, who found that the rise-time ruled out PISN models for two slowly-declining SLSNe, PTF12dam and PS1-11ap.

\subsection{A Nickel-powered precursor?}\label{sec:ni}

\begin{figure}
\includegraphics[width=\columnwidth,angle=0]{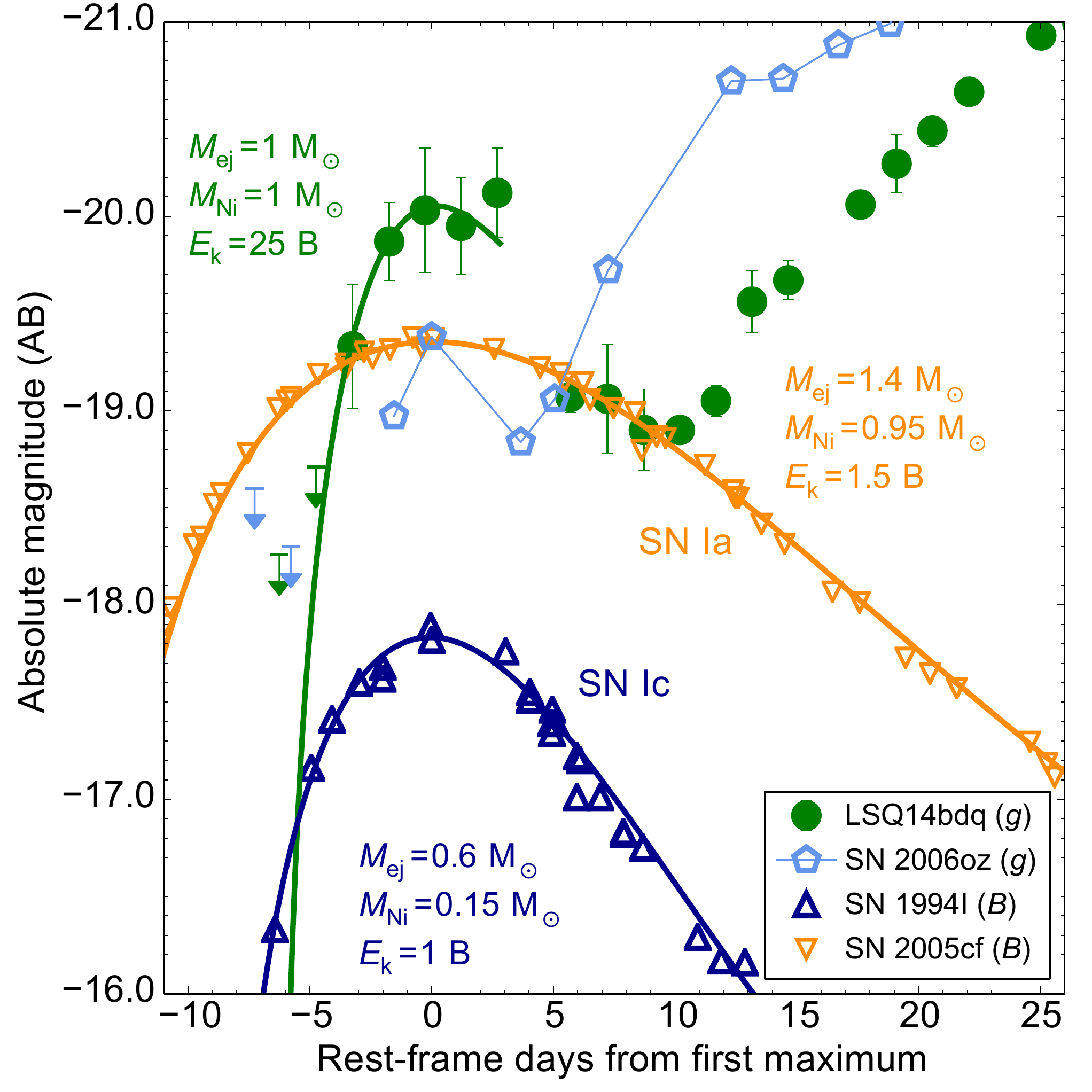}
\caption{Nickel-powered model for the precursor peak. The early light curve of LSQ14bdq is similar to SN 1994I (Type Ic) but is much brighter, whereas it is much narrower than the Type Ia SN 2005cf. The combination of high luminosity and very short rise-time rules out a physically-plausible \Ni-powered SN for this peak (see Sect.~\ref{sec:ni}).
}
\label{fig:nickel}
\end{figure}

The first scenario we investigate for the early peak is an initially normal SN, powered by the radioactive decay of \Ni, before the mechanism powering the super-luminous second peak kicks in. We compare our early light curve to \Ni-powered SNe (core-collapse and thermonuclear) in Fig.~\ref{fig:nickel}, choosing filters with similar effective wavelengths. As the late-time spectra of Type Ic SLSNe closely resemble SNe Ic, we first compare to SN 1994I, a well-observed object with a narrow light curve, suggesting \Mej$\,\lesssim1\,$\M~\citep{ric1996}. The width of the LSQ14bdq peak is slightly narrower, but comparable to SN 1994I. However, it is 2.3 magnitudes brighter. Using the \citet{arn1982} model, as implemented by \citet{ins2013}, we find that matching the photometry\footnote{Assuming a blackbody SED, we apply synthetic photometry to the model using \textsc{Pysynphot} (http://stsdas.stsci.edu/pysynphot/).} requires an almost pure \Ni~ejecta of $\simeq1\,$\M, which is difficult to produce with core-collapse SNe. Type Ia SNe produce $\approx0.5$--1\,\M~of nickel, but comparing with SN 2005cf \citep{pas2007} shows that SNe Ia have light curves that are too broad. The fast rise of LSQ14bdq would necessitate a very large explosion energy, even for the lowest possible ejecta mass, \Mej$\,=\,$\Mni. A fit shows that we require $E_{\rm k}=25\,$B, where 1\,B\,(Bethe)\,$=10^{51}\,$erg. Because complete burning of 1\,\M~of carbon/oxygen to nickel liberates only $\sim1$\,B, an additional energy source is required. Thus, if the early peak is `a supernova in itself', it cannot be a normal \Ni-powered event.

\subsection{Shock cooling with a central engine}\label{sec:cool}

\begin{figure}
\includegraphics[width=\columnwidth,angle=0]{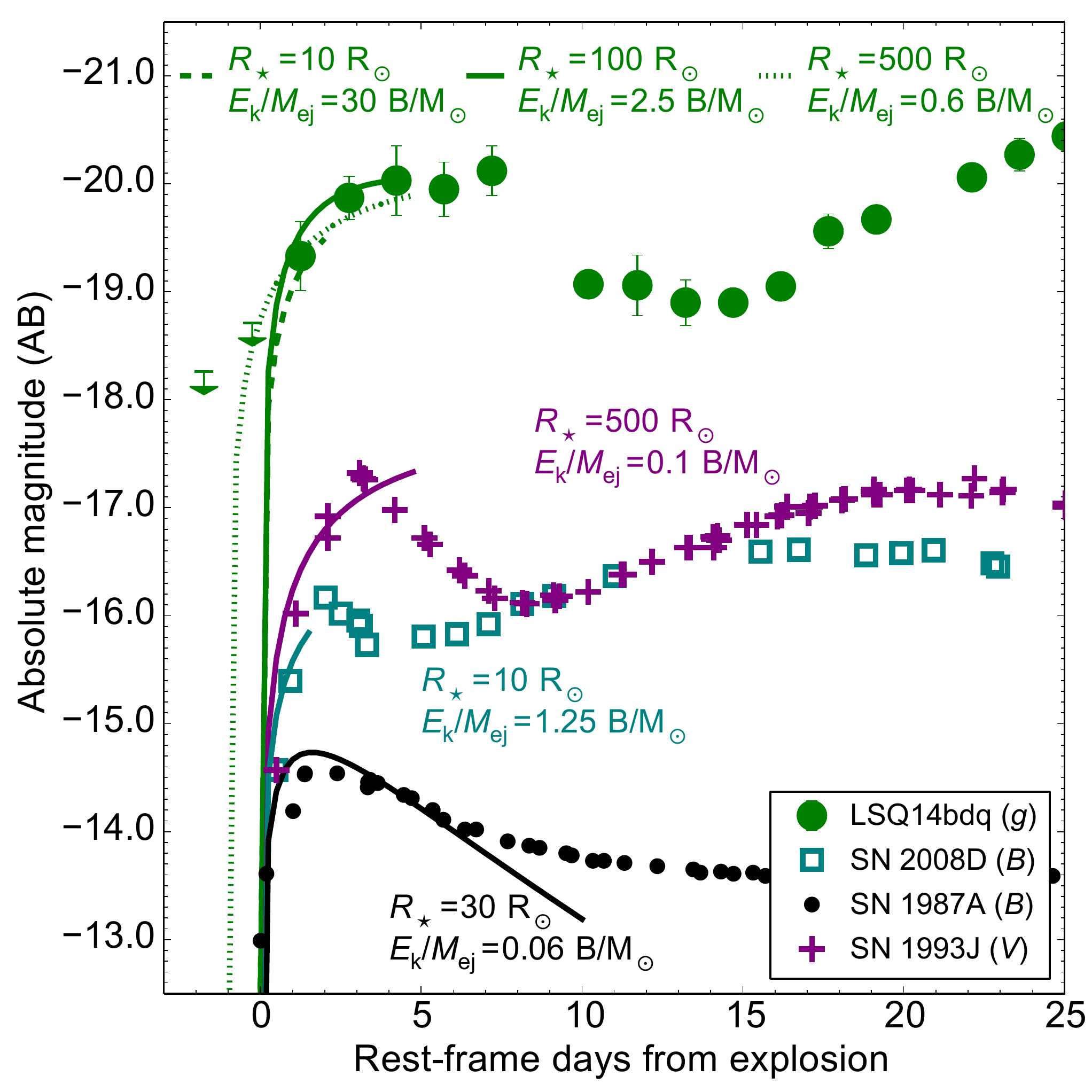}
\caption{Post-shock cooling models for LSQ14bdq and other double-peaked SNe. We show both compact and extended models. Mass and energy can be inferred from the fits together with equations \ref{eq:M} and \ref{eq:E}. More extended progenitors can reproduce the peak brightness with lower \Mej~and \E.
}
\label{fig:cool}
\end{figure}

In Fig.~\ref{fig:cool}, we show an alternative scenario, comparing the early emission from LSQ14bdq to other SNe with two observed peaks. The basic light curve morphology of SNe 1987A, 1993J and 2008D has been interpreted as an initial cooling phase, releasing heat deposited by the shock wave, before a second peak is driven by delayed heating from \Ni~\citep{shig1990,woo1994,mod2009}. The light curve of LSQ14bdq is qualitatively similar to these objects, suggesting that the light curve could be explained by a shock cooling phase, followed by internal reheating. While bolometric luminosity in the cooling phase should decline monotonically, single-filter light curves show maxima as the peak of the SED moves into and out of the optical.

 We fit the early rise using analytic approximations from \citet{rab2011}, giving the parameters in Fig.~\ref{fig:cool}. We use their blue supergiant (radiative envelope), red supergiant (convective envelope) and Wolf-Rayet models for SN 1987A, 1993J and 2008D, respectively. The progenitor radius, \Rs, determines the slope and duration of the rise, while the luminosity scale is set by both \Rs~and by the explosion energy per unit mass, \EM. The values of \Rs~and \EM~used to fit the literature objects are in line with previous estimates. As noted by \citet{rab2011}, the model assumptions begin to break down after a few days; we end the simulations at the cut-off time prescribed by their equation 17. The model for SN 1993J is still rising slowly as it goes through the peak. The discrepancy with observations may be due to the very simple density profile assumed in the model. Detailed models of SN 1993J have had $\lesssim 0.1$\M~in the extended envelope, with most of its mass in the core.

To model LSQ14bdq, we set \EM$\,=C$\,B/\M, where $C$ is arbitrary. To try to break the degeneracy between \E~and \Mej, we assume that the diffusion time during the early peak is the same as in our central-engine fit to the main peak. We neglect late-time kinetic energy input from the magnetar, $\approx10^{51}$\,erg for our model, because this is small compared to the initial energy found for the shock cooling, as will be seen below. We then have
\begin{equation}\label{eq:diff}
\tau_m = \frac{1.05}{\left(13.7 c \right)^{1/2}} \kappa^{1/2} \left( \frac{M_{\rm ej}^3}{E_{\rm k}} \right)^{1/4} = 90\,{\rm days}.
\end{equation}
For $\kappa=0.2\,{\rm cm}^2\,{\rm g}^{-1}$, this leads to:
\begin{equation}\label{eq:M}
M_{\rm ej} = 40.1\,C^{1/2}\,{\rm M}_{\odot};
\end{equation}
\begin{equation}\label{eq:E}
E_{\rm k} = 40.1\,C^{3/2}\,{\rm B}.
\end{equation}
The uncertainty in $\tau_m$ (taking the range where $\chi^2<2 \chi^2_{\rm min}$) is $\sim30\%$, meaning \M~and \E~are constrained to within a factor of 2.

Fig.~\ref{fig:cool} shows models for 3 progenitors: a Wolf-Rayet with \Rs\,$=10$\,\R, and extended stars with \Rs\,$=100$,\,500\,\R~(extended models are insensitive to the choice of radiative/convective envelope). For the compact model, we derive \Mej$\,\approx270\,$\M; depending on the precise mass, the implied progenitor should either explode as a PISN or collapse totally (hence invisibly) to a black hole \citep{heg2002}, neither being consistent with the light curve. The inferred energy, \E\,$>10^{54}$\,erg, is also unrealistic.

An extended envelope \citep[or wind;][]{ofek2010} is therefore a requirement in this scenario. The 100\,\R~model requires \Mej$\,\approx 60\,$\M~and \E\,$\approx150$\,B. This energy is greater than the canonical neutron star gravitational binding energy of $10^{53}$\,erg (of which $\sim1\%$ is normally accessible to power the explosion). The energy released in black-hole-formation is higher than for neutron stars and could meet the requirement, if it could couple to the ejecta. A possible mechanism is an accretion disk, such as in the collapsar model \citep{woo1993} of gamma-ray bursts (GRBs). In this case the engine would be black hole accretion \citep{dex2013} rather than a magnetar. This accretion engine has a characteristic power law, $L \propto t^{-n}$, similar to the magnetar, with $n_{\rm mag}=2$ and $n_{\rm acc}=5/3$. Therefore we would expect a similar $\tau_m$, and that equations \ref{eq:M} and \ref{eq:E} would still hold.

The final model shown is for \Rs\,$=500$\,\R. The inferred mass is \Mej\,$\approx 30\,$\M, with \E\,$\approx 20$\,B, which may also favour a black hole engine over a neutron star (it is similar to the kinetic energy in GRB-SNe), but not so definitively as in the more compact models. The radius is very large for a hydrogen-free star, but similar to SN 1993J, which had only a very diffuse hydrogen envelope, and by maximum light had evolved to resemble a SN Ib. For this model, the velocity, $v\sim \sqrt{10 E_{\rm k}/(3 M_{\rm ej})} = 10000$\,\kms, is in good agreement with the observed spectrum.

\subsection{CSM interaction}\label{sec:csm}


An alternative scenario to consider is that the main peak arises from CSM interaction on scales of $\sim 10^4$\,\R, as has been suggested for other SLSNe \citep[e.g.][]{cha2013}. CSM fits for the main peak (Fig.~\ref{fig:lc}) require \EM\,$\sim 0.2$\,B/\M, lower than any of the shock cooling models shown in Fig.~4. This is fairly inflexible for the CSM model, as \E~and \Mej~are the two strongest drivers of the peak luminosity -- e.g.~models with \EM\,$\sim 0.4$\,B/\M~are too bright by about 0.5 dex (and rise too quickly). To reproduce the early emission with shock cooling and \EM\,$\sim$\,0.2\,B/\M, we would need initial radius \Rs$\,\sim2000$\,\R. This is uncomfortably large for any reasonable progenitor. However models have been proposed in which the cooling phase arises from shock breakout in an inner region of dense CSM rather than the progenitor envelope \citep{ofek2010}; this may be a viable explanation for LSQ14bdq, but would require a novel CSM structure to get two distinct peaks.

A model for SN 2006oz was put forward by \citet{mor2012}, in which a single CSM interaction produces a double-peaked light curve due to a sudden increase in CSM ionisation (and hence opacity) in the collision. However, this model does not specify the source of the early emission before the collision, which for LSQ14bdq we have shown rises too steeply to be explained as a conventional SN. Multi-peaked light curves from interaction could also arise within the pulsational pair-instability model \citep{woo2007}. Here, multiple shells are ejected with $v \sim 3000 - 5000$\,\kms, and could produce distinct interaction events with a range of luminosities and timescales. Suppose SN ejecta collide with an inner shell at the beginning of our observations, and the resulting merged ejecta/shell then hit an outer shell fifteen days later, generating the second peak. For a shell separation $R_{\rm CSM} \sim 10^{15}$\,cm (from our CSM fit and the \citealt{woo2007} models), this would require $v_{\rm ej+inner} \sim 10^4$\,\kms, which is similar to the observed line widths. This may provide a reasonable alternative to the shock cooling scenario, although the massive outer shell must also be accelerated to avoid showing narrow spectral lines. We note that the \citet{woo2007} models are much redder than our observations. Hence further detailed modelling is needed to assess the viability of this scenario for LSQ14bdq.

\section{Discussion and conclusions}

The detection of a double-peaked light curve provides a new opportunity to constrain the physics powering SLSNe Ic. 
We find that the early peak is not likely to be a normal \Ni-powered SN. We also disfavour the CSM interaction model, since a consistent physical scenario seems to require a helium star with an extremely large radius \Rs\,$>2000$\,\R, within a more extended (and hydrogen poor) CSM of $R_{\rm CSM}\gtrsim10^4$\,\R. Pulsational pair-instability models remain a possibility, but do not yet quantitatively reproduce the observed data. 

We propose that the initial peak could arise from post-shock cooling, and provide a simple physical interpretation consistent with the main light curve. The first peak is itself remarkably bright ($M_g = -20.0$), suggesting a large stellar radius or high explosion energy, or both. The broad width of the second peak implies a large ejected mass, and can be powered by a central engine. We find good fits for an explosion with $E_{\rm k} \sim 20$\,B, in a star with $R_{\star}\sim500$\,\R, that ejected a mass of $\sim$30\,\M. This energy may favour a black hole accretion engine \citep{dex2013} rather than a magnetar, and is similar to kinetic energies seen in long GRBs. 

The extended radius is surprising, and argues against a compact Wolf-Rayet progenitor ($R\lesssim$10\,\R) since it would imply unrealistic ejecta mass and explosion energy. Although such extended He stars are not known in the local Universe, the fact that SLSNe may be confined to very low metallicity galaxies and are intrinsically rare ($\sim$1 in 10,000 massive star deaths) may explain the lack of known counterparts.

Further, early observations of SLSNe will determine whether or not double-peaked light curves are common. 
Theoretical stellar evolution models with binarity and rotation should be explored for a viable progenitor.
\\

Based on data from ESO as part of PESSTO (188.D-3003, 191.D-0935), 
2.2-m MPG telescope (CN2014B-102, GROND DFG grant HA 1850/28-1).
PS1 is supported by NASA Grants NNX12AR65G and NNX14AM74G from NEO Observation Program. We acknowledge: EU/FP7-ERC  Grants [291222, 307260, 320360] (SJS, AVG, MS) and FP7 grant agreement n. 267251 (NER); CONICYT-Chile FONDECYT grants 3140566, 3140534, Basal-CATA PFB-06/2007, and the Millennium Science Initiative grant IC120009 to MAS (LG, SS); PRIN-INAF 2014 project ÒTransient Universe: unveiling new types of stellar explosions with PESSTOÓ (SB).

\begin{table*}
\caption{Observed photometry of LSQ14bdq}
\label{tab:phot}
\begin{center}
\begin{tabular}{cccccccc}
\hline
\hline
Date & MJD & Phase* & $g$ & $r$ & $i$ & $z$ & Telescope\\
\hline
02-03-2014	&	56718.0	&	-66.2	&	&	$>$22.85		&		&		&	LSQ	\\
04-03-2014	&	56720.0	&	-64.7	&	&	$>$22.40		&		&		&	LSQ	\\
06-03-2014	&	56722.0	&	-63.2	&	&	21.78 (0.32)	&		&		&	LSQ	\\
08-03-2014	&	56724.0	&	-61.7	&	&	21.24 (0.20)	&		&		&	LSQ	\\
10-03-2014	&	56726.0	&	-60.2	&	&	21.08 (0.32)	&		&		&	LSQ	\\
12-03-2014	&	56728.0	&	-58.7	&	&	21.16 (0.25)	&		&		&	LSQ	\\
14-03-2014	&	56730.0	&	-57.3	&	&	20.99 (0.23)	&		&		&	LSQ	\\
18-03-2014	&	56734.0	&	-54.3	&	&	22.04 (0.08)	&		&		&	LSQ	\\
20-03-2014	&	56736.1	&	-52.7	&	&	22.05 (0.28)	&		&		&	LSQ	\\
22-03-2014	&	56738.1	&	-51.2	&	&	22.21 (0.21)	&		&		&	LSQ	\\
24-03-2014	&	56740.1	&	-49.7	&	&	22.21 (0.03)	&		&		&	LSQ	\\
26-03-2014	&	56742.1	&	-48.3	&	&	22.06 (0.08)	&		&		&	LSQ	\\
28-03-2014	&	56744.1	&	-46.8	&	&	21.55 (0.16)	&		&		&	LSQ	\\
30-03-2014	&	56746.1	&	-45.3	&	&	21.44 (0.10)	&		&		&	LSQ	\\
03-04-2014	&	56750.1	&	-42.3	&	&	21.05 (0.02)	&		&		&	LSQ	\\
05-04-2014	&	56752.1	&	-40.9	&	&	20.84 (0.15)	&		&		&	LSQ	\\
07-04-2014	&	56754.0	&	-39.4	&	&	20.67 (0.08)	&		&		&	LSQ	\\
09-04-2014	&	56756.1	&	-37.9	&	&	20.47 (0.03)	&		&		&	LSQ	\\
13-04-2014	&	56760.1	&	-34.9	&	&	20.18 (0.06)	&		&		&	LSQ	\\
15-04-2014	&	56762.1	&	-33.4	&	&	20.12 (0.03)	&		&		&	LSQ	\\
17-04-2014	&	56764.0	&	-31.9	&	&	19.93 (0.05)	&		&		&	LSQ	\\
19-04-2014	&	56766.0	&	-30.5	&	&	19.88 (0.02)	&		&		&	LSQ	\\
21-04-2014	&	56768.0	&	-28.9	&	&	19.89 (0.07)	&		&		&	LSQ	\\
23-04-2014	&	56770.0	&	-27.5	&	&	19.74 (0.01)	&		&		&	LSQ	\\
25-04-2014	&	56772.1	&	-25.9	&	&	19.71 (0.05)	&		&		&	LSQ	\\
28-04-2014	&	56775.5	&	-23.4	&	&	19.64 (0.19)	&		&		&	PS1	\\
06-05-2014	&	56783.1	&	-17.8	&	19.55 (0.09)	&	19.35 (0.12)	&	19.45 (0.05)	&		&	NTT	\\
07-05-2014	&	56784.7	&	-16.6	&	19.29 (0.11)	&	19.31 (0.20)	&		&		&	LCOGT	\\
12-05-2014	&	56789.9	&	-12.7	&	19.27 (0.04)	&	19.26 (0.12)	&	19.18 (0.06)	&	19.48 (0.12)	&	LT	\\
20-05-2014	&	56797.9	&	-6.8	&	19.27 (0.04)	&	19.16 (0.02)	&	19.16 (0.06)	&	19.40 (0.04)	&	LT	\\
23-11-2014	&	56984.2	&	131.8	&	&	21.15 (0.23)	&		&		&	LSQ	\\
23-11-2014	&	56984.3	&	131.8	&	&	20.99 (0.10)	&	20.89 (0.06)	&	&	NTT	\\
27-11-2014	&	56988.2	&	134.7	&	&	21.05 (0.24)	&		&		&	LSQ	\\
28-11-2014	&	56989.2	&	135.5	&	&	21.36 (0.30)	&		&		&	LSQ	\\
18-12-2014	&	57009.2	&	150.3	&	&	21.27 (0.33)	&		&		&	LSQ	\\
20-12-2014	&	57011.5	&	152.0	&	&	21.51 (0.24)	&		&		&	PS1	\\
23-11-2014	&	57024.2	&	161.5	&	&	21.78 (0.38)	&		&		&	LSQ	\\
02-01-2015	&	57033.2	&	168.2	&	&	21.65 (0.23)	&		&		&	LSQ	\\
20-01-2015	&	57042.0	&	174.3	&	&	22.16 (0.20)	&		&		&	PS1	\\
12-02-2015	&	57065.3	&	192.0	&	23.29 (0.11)	&	22.41 (0.04)	&	22.30 (0.08)	&	22.00 (0.17)	&	GROND	\\
26-02-2015	&	57079.2	&	202.4	&	23.74 (0.14)	&	22.89 (0.13)	&	22.82 (0.31)	&	22.79 (0.41)	&	NTT	\\
13-03-2015	&	57094.1	&	213.5	&	24.35 (0.13)	&	23.43 (0.16)	&	22.87 (0.24)	&	22.45 (0.32)	&	NTT	\\
\hline
\multicolumn{3}{l}{Host (2012-2013 stack)}	&	&	$>$24.08		&		&		&	LSQ	\\
\hline
\hline
\multicolumn{3}{c}{$K$-corrections$^{\dagger}$}	& $K_{g \rightarrow u}$	&	$K_{r \rightarrow g}$	&	$K_{i \rightarrow r}$	&	$K_{z \rightarrow i}$	&	\\
\hline
\multicolumn{3}{l}{Phase \textless 0 (LSQ14bdq spectrum, $-$19d)}	&		$-$0.30	&	$-$0.32	&	$-$0.49	&	$-$0.38	&	 \\
\multicolumn{3}{l}{Phase \textgreater 0 (PTF12dam spectrum, +171d)}	&	$-$0.22	&	$-$0.16	&	$-$0.18	&	$-$0.20	&	 \\
\hline
\hline
\multicolumn{3}{c}{Near-infrared (Vega system)}	& J	& H	&	K 	& 	\\
\hline
12-02-2015	&	57065.3	&	192.0	&	20.70 (0.17)	&	$>$20.03	&	$>$18.39 	&	&	GROND	\\
\hline

\end{tabular}
\end{center}
*Rest-frame days relative to the estimated date of $r$-band maximum, MJD=56807.\\
$\dagger$Defined by $m = M + \mu + K$
\end{table*}

\bibliographystyle{apj}

\end{document}